\documentclass[a4paper,11pt]{article}
\pdfoutput=1 

\usepackage{jinstpub} 

\usepackage{lineno}
\usepackage{graphicx}

\title{\boldmath Upgrade of Belle II Vertex Detector with CMOS Pixel Technology}

\author[a,1]{M. Schwickardi\note{Corresponding author.}}
\author[b]{, M.~Babeluk}
\author[c]{, M.~Barbero} 
\author[d]{, J.~Baudot} 
\author[b]{, T.~Bergauer} 
\author[d]{, G.~Bertolone} 
\author[e,f]{, S.~Bettarini} 
\author[f]{, F.~Bosi} 
\author[c]{, P.~Breugnon} 
\author[a]{, Y.~Buch} 
\author[e,f]{, G.~Casarosa}
\author[d]{, G. Dujany}
\author[d]{, C.~Finck} 
\author[e,f]{, F.~Forti}
\author[a]{, A.~Frey} 
\author[d]{, A.~Himmi} 
\author[b]{, C.~Irmler} 
\author[d]{, A.~Kumar} 
\author[g]{, C.~Marinas} 
\author[f]{, M.~Massa} 
\author[e,f]{, L.~Massaccesi} 
\author[g]{, J.~Mazzora~de~Cos}
\author[f]{, M.~Minuti}
\author[e,f]{, S.~Mondal} 
\author[c]{, P.~Pangaud} 
\author[d]{, H.~Pham} 
\author[d]{, I. Ripp-Baudot}
\author[e,f]{, G.~Rizzo}
\author[a]{, B.~Schwenker} 
\author[d]{, and I. Valin}

\affiliation[a]{Georg-August Universität, 37077 Göttingen, Germany}
\affiliation[b]{Institute of High Energy Physics, Austrian Academy of Sciences, 1050 Vienna, Austria}
\affiliation[c]{Aix Marseille Universit\'{e}, CNRS/IN2P3, CPPM, 13288 Marseille, France}
\affiliation[d]{Universit\'{e} de Strasbourg, CNRS, IPHC, UMR 7178, 67037 Strasbourg, France}
\affiliation[e]{Dipartimento di Fisica, Universit\`{a} di Pisa, I-56127 Pisa, Italy}
\affiliation[f]{INFN Sezione di Pisa, I-56127 Pisa, Italy}
\affiliation[g]{Instituto de Fisica Corpuscular, IFIC, CSIC-UV, Paterna 46980, Spain}

\emailAdd{marike.schwickardi@uni-goettingen.de}

\abstract{The Belle II experiment at KEK in Japan considers upgrading its vertex detector system to address the challenges posed by high background levels caused by the increased luminosity of the SuperKEKB collider. One proposal for upgrading the vertex detector aims to install a 5-layer all monolithic pixel vertex detector based on fully depleted CMOS sensors in 2027. The new system will use the OBELIX MAPS chips to improve background robustness and reduce occupancy levels through small and fast pixels. This causes better track finding, especially for low transverse momenta tracks. This text will focus on the predecessor of the OBELIX sensor, the TJ-Monopix2, presenting laboratory and test beam results on pixel response, efficiency, and spatial resolution.}


\collaboration[c]{on Behalf of the Belle II VTX Upgrade Group}

\proceeding{24$^{\text{th}}$ international
Workshop on Radiation Imaging Detectors\\
  25-29 JUNE 2023\\
  Oslo}

\begin{document}
\maketitle
\flushbottom

\section{Introduction}
\label{sec:intro}

The Belle\;II experiment \cite{belle} is a high-energy physics experiment designed as an all-purpose full solid angle particle detector. It is located at the SuperKEKB accelerator facility \cite{Akai_2018} in Tsukuba, Japan. The Belle\;II experiment operates at a centre of mass energy of $\sqrt{s} = 10.58$\;GeV, corresponding to the mass of the $\Upsilon(4S)$ particle. This choice of energy is optimal for the production of $B\Bar{B}$ pairs just above the production threshold. The SuperKEKB accelerator, which collides 7\;GeV electrons with 4\;GeV positrons, is an asymmetric collider. This means that the energy of the electron beam is significantly higher than that of the positron beam, causing the produced particles to be boosted along the z-direction. This enables time-resolved measurements on the $B\Bar{B}$ decays and allows for a more precise reconstruction of the momenta, given a highly granular vertex detector.

The aim is to take data at an instantaneous luminosity of $6\times10^{35} \text{cm}^{-2}\text{s}^{-1}$ provided by the SuperKEKB collider. The beam conditions to reach such luminosity levels generate a large rate of background particles in the inner detection layers of Belle II. The background hit rate is 120~MHz/cm$^2$ in the innermost layer at maximum, translating into radiation levels of 10 Mrad/year ionising dose and fluences reaching $5\times10^{13}~\text{n}_{\text{eq}}\text{cm}^{-2}/\text{year}$ \cite{snowmass}.
The upcoming SuperKEKB upgrade in 2027/28 might necessitate a redesign of the interaction region to reach the luminosity goal of $6\times10^{35}\text{cm}^{-2}\text{s}^{-1}$. This presents a strategic opportunity  to replace the existing vertex detector (VXD) with the incorporation of depleted CMOS MAPS (Complementary Metal-Oxide-Semiconductor Monolithic Active Pixel Sensors) technology. The objective of this upgraded vertex detector is to substantially elevate tracking and vertexing capabilities within the demanding high luminosity setting.

\section{VTX Proposal}

The current Vertex Detector (VXD) of the Belle\;II experiment consists of two sub-detector systems: the Pixel Detector (PXD) and the Silicon Vertex Detector (SVD). The PXD  uses DEPFET-based \cite{KEMMER1987365} pixel technology with thin sensors (75\;\textmu m thickness) and a small pixel pitch (50-75\;\textmu m) as described in \cite{Avella_2014}. Its readout employs a rolling shutter, resulting in a rather long integration time of 20\;\textmu s. The SVD \cite{Adamczyk_2022} comprises four layers of double-sided strip detectors. It exhibits an excellent cluster time resolution of 3\;ns \cite{ZANI2022166952}. The spatial resolution is constrained to 10-25\;\textmu m due to the strip pitch \cite{ZANI2022166952}. However, the space granularity limits the occupancy to around 4\% due to the strips' length. The primary objectives of the VXD are accurate decay vertex estimation and identification of very low momentum tracks. To maintain a low material budget, the VXD was constructed with a material budget equivalent to about 3\% of the radiation length ($X_0$).

\begin{figure}
    \begin{minipage}[b]{0.49\textwidth}
        \centering
        \includegraphics[width=1\textwidth]{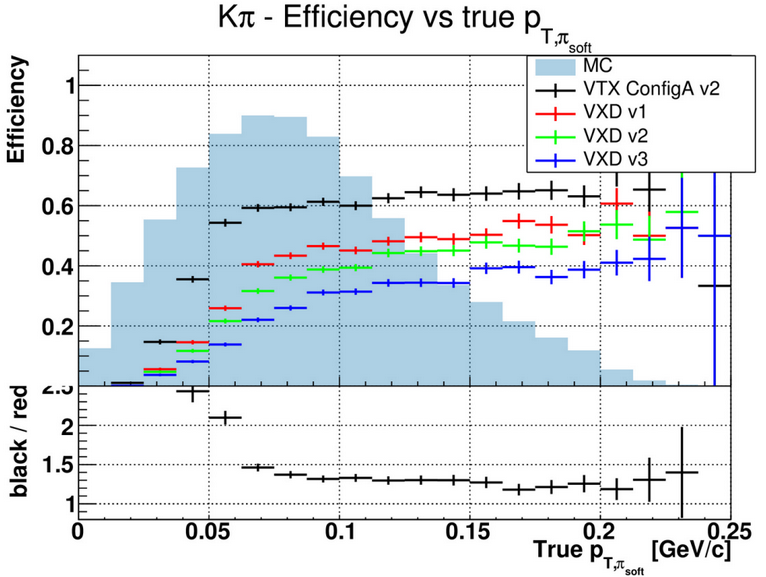}
        \caption{Comparing Soft Pion Signal Efficiency in the Low p$_\text{t}$ Range for VTX (black) and VXD under three background scenarios (red - best, green - average, blue - worst) via Monte Carlo simulation.}
        \label{fig:soft_pion}
    \end{minipage}
    \hspace{0.15cm}
    \begin{minipage}[b]{0.49\textwidth}
        \centering
        \includegraphics[width=1\textwidth]{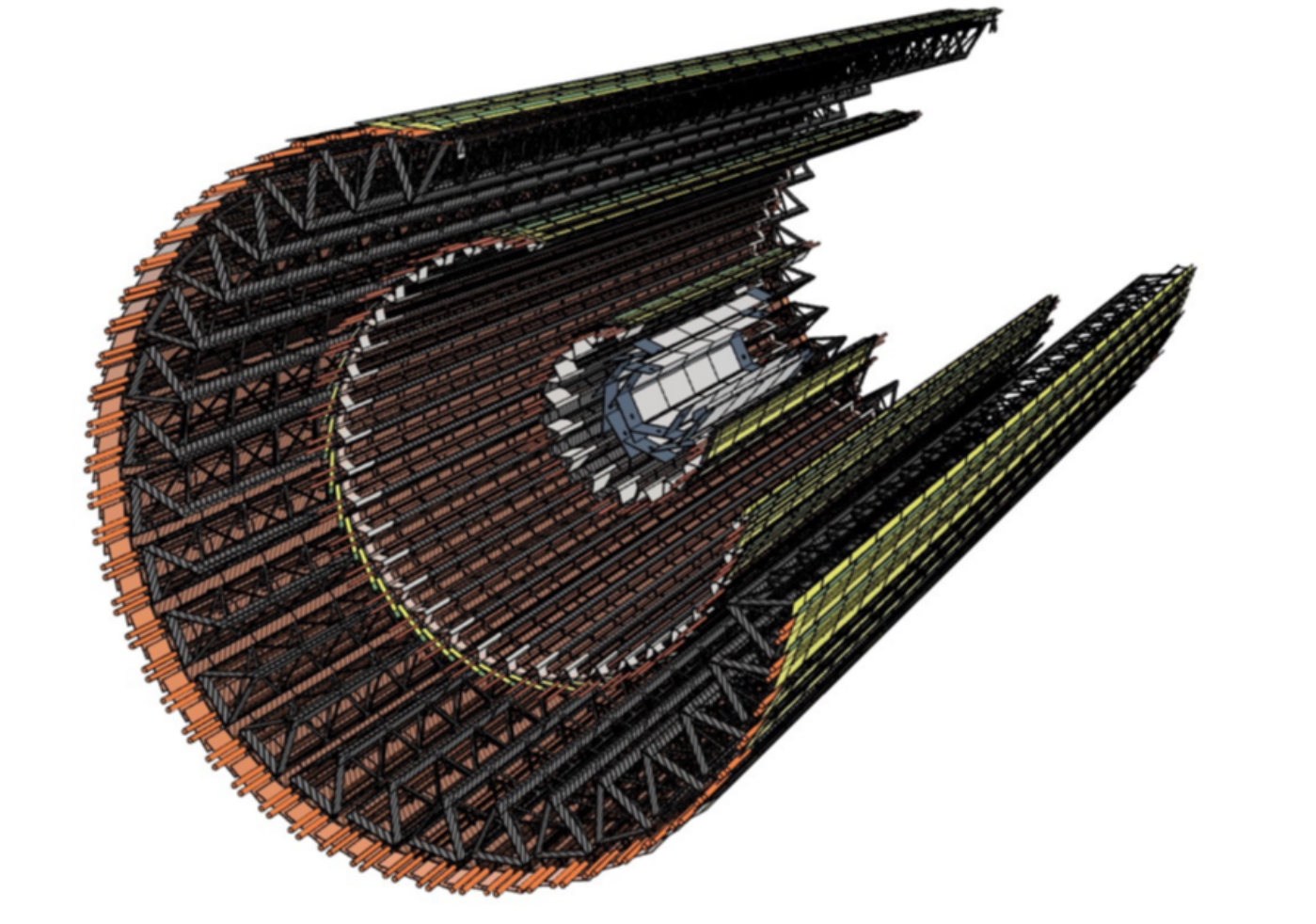}
        \caption{Illustration showcasing the advanced Belle II Vertex Detector VTX, distinguished by its utilization of five layers of depleted CMOS technology, as elaborated upon in reference \cite{BABELUK2023168015}.}
        \label{fig:vtx_cad}
    \end{minipage}%
\end{figure}

Currently, the PXD modules exhibit a maximum occupancy just below 1\%. However, it is important to note that performance degradation is possible at higher occupancy levels of 2-3\%. To assess potential impacts, three scenarios have been developed for the purpose of background extrapolation in the context of design luminosity. Under each scenario, simulations of background hits are executed, followed by an evaluation of their influence on the detector performance. These simulations consider the best-case scenario (1), an average use-case scenario (2), and the worst-case scenario (3). The extrapolations indicate that the sensors may approach the limits of the current detector in high luminosity environments. Even at a PXD occupancy of 1\%, PXD hits cannot be utilised for the track reconstruction of low transverse momentum (low pt) tracks, as depicted in Figure \ref{fig:soft_pion}. The graph presents the performance predictions of the VXD, specifically regarding its ability to efficiently detect soft pion signals within the low pt (transverse momentum) range. The Monte Carlo (MC) spectrum is shown and a low soft pion signal efficiency is indicated for all three background scenarios. 

The  upgrade presents an opportunity to replace the current vertex detector by implementing a 5 straight layer depleted CMOS MAPS Detector called VTX, as shown in Figure \ref{fig:vtx_cad}. The figure shows a CAD drawing of the central beam pipe with the electron and positron beam indicated, surrounded by the two innermost layers, which make up the iVTX in the centre and the three outer layers that make up the oVTX. One of the primary goals of this upgrade is to reduce the material budget to approximately 2.5\%~$X_0$ to minimise multiple Coulomb scattering and improve the precision of particle trajectory measurements. The incorporation of depleted CMOS MAPS will also lead to an increase in space-time granularity, allowing for more precise tracking of particle interactions. To meet the requirements of this upgrade, the new vertex detector must be robust against the inner layer background, as well as handle a hit-rate of up to 120~MHz/cm\textsuperscript{2}.

Another requirement on the sensors is that the resolution of the detector has to be better than 15~\textmu m to achieve high-precision measurements of particle positions. High efficiency is another essential factor to ensure that a large number of particle tracks are successfully detected. The vertex detector is also expected to withstand ionising doses of approximately 1~Mrad per year and handle a Non-Ionizing Energy Loss (NIEL) of $5 \times 10^{13}$~n$_\text{eq}$/cm\textsuperscript{2} per year.

As previously mentioned, the VTX is designed with a configuration consisting of five layers. The detector employs a ladder/stave setup, with the initial two layers, L1 and L2, forming the inner vertex detector (iVTX). These specific layers, L1 and L2, are constructed using all-silicon ladders and employ an air-cooling mechanism, resulting in a material budget of around 0.1\% $X_0$ per layer. Conversely, the outer vertex detector (oVTX) is composed of three layers: L3, L4, and L5. For layers L3 and L4, a material budget ranging from approximately 0.3\% to 0.5\% $X_0$ is achieved by implementing a support frame made of carbon fibre. In the case of layer L5, a material budget of roughly 0.8\% $X_0$ is anticipated. This particular layer incorporates a cooling plate equipped with water cooling to efficiently dissipate heat. This configuration holds the promise of enhancing the precision and effectiveness of particle tracking within a high luminosity environment. The performance of the VTX is evaluated through simulations similar to the ones performed for the VXD. When Full Tracking is considered (which involves the fusion of vertex tracking with the CDC - Central Drift Chamber), the VTX outperforms the VXD in terms of tracking efficiency. Notably, the soft pion signal efficiency is significantly impacted by the VTX performance, as shown in Figure \ref{fig:soft_pion}.

\section{TJ-Monopix2}
The VTX detector will be equipped with the OBELIX chip, which is currently under development and is planned for submission in autumn 2023. The sensor matrix is based on the TJ-Monopix2 \cite{BESPIN2020164460,MOUSTAKAS2019604}, initially developed for the ATLAS ITk outer layer requirements, where the pixel matrix will basically be copied and a trigger adaptation will be added in the periphery.
The TJ-Monopix2 is fabricated using the TowerJazz (TJ) 180\;nm process, and the chip has a size of 2\;$\times$\;2\;cm\textsuperscript{2}, comprising a pixel matrix of 512\;$\times$\;512\;pixels with a pixel pitch of 33.04\;\textmu m.
The design's specifications include an expected minimum threshold of approximately 100\;$e^-$ from simulations, with a tuned threshold dispersion of 5-10\;$e^-$. It aims to achieve over 97\% efficiency for MIPs at 10\textsuperscript{15}\;n$_\text{eq}$/cm\textsuperscript{2} and a noise level of around 5\;$e^-$. The chip is expected to remain fully efficient for hit rates of 120\;MHz/cm\textsuperscript{2}, however the periphery of TJ-Monopix2 is limited to lower rates.

For exploration purposes, the current sensors are equipped with four different Front-End (FE) flavours, each having distinct characteristics in terms of the pre-amplifier, the coupling to the collection node, and biasing.
Firstly, there are two types of DC-coupled FEs: Normal and Cascode.  
Additionally, there are HV (High Voltage) and HV Cascode FE variants that are AC-coupled to the charge collection electrode. This AC-coupling feature allows for higher bias voltages. The Cascode and Non-Cascode versions of these FE flavours differ only by the presence of one transistor that is specifically designed to enhance the gain.

To comprehensively characterise  the sensor, a variety of laboratory tests are employed. One of these methods is the internal injection test  where a known amount of charges is introduced at the input of the pre-amplifier of the system. The output generated is then quantified using the Time over Threshold (ToT), which is measured in units of 25 nanoseconds with a 7-bit counter. Further details on the injection testing can be found in \cite{Bespin_2023}.
Furthermore, it is possible to conduct an absolute calibration by taking measurements using a Fe$^{55}$ source. This source exhibits a distinct X-ray peak with an energy of 5.9 keV which serves as a reference point for calibrating the injected charge in terms of electron-hole pairs. 
These tests are conducted across several locations, including University of Bonn, INFN Pisa, HEPHY Vienna, CPPM Marseilles, IPHC Strasbourg, and University of Göttingen.

\begin{figure}
    \begin{minipage}[b]{0.49\textwidth}
\centering
\includegraphics[width=1\textwidth,  page=2 ,trim=0.2cm 0.2cm 1.1cm .7cm, clip,]{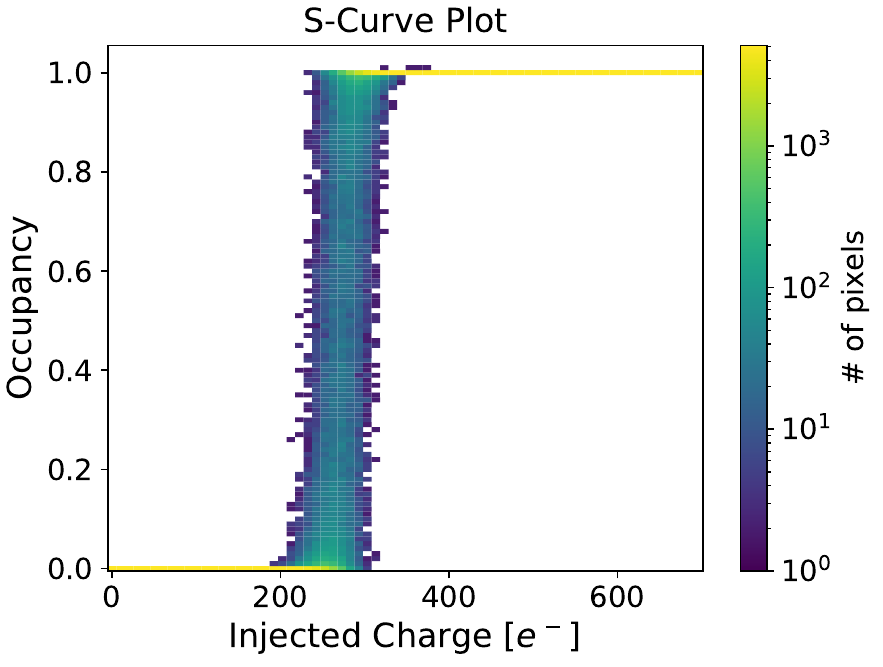}
        \caption{S-curve measurement of TJ-Monopix2 DC coupled FE with mean and dispersion for tuned settings.}
        \label{fig:s-curve}
    \end{minipage}%
    \hspace{0.15cm}
    \begin{minipage}[b]{0.49\textwidth}
\centering
\includegraphics[width=1\textwidth, ,trim=0.2cm 0.0cm 0.2cm .7cm, clip, page=4]{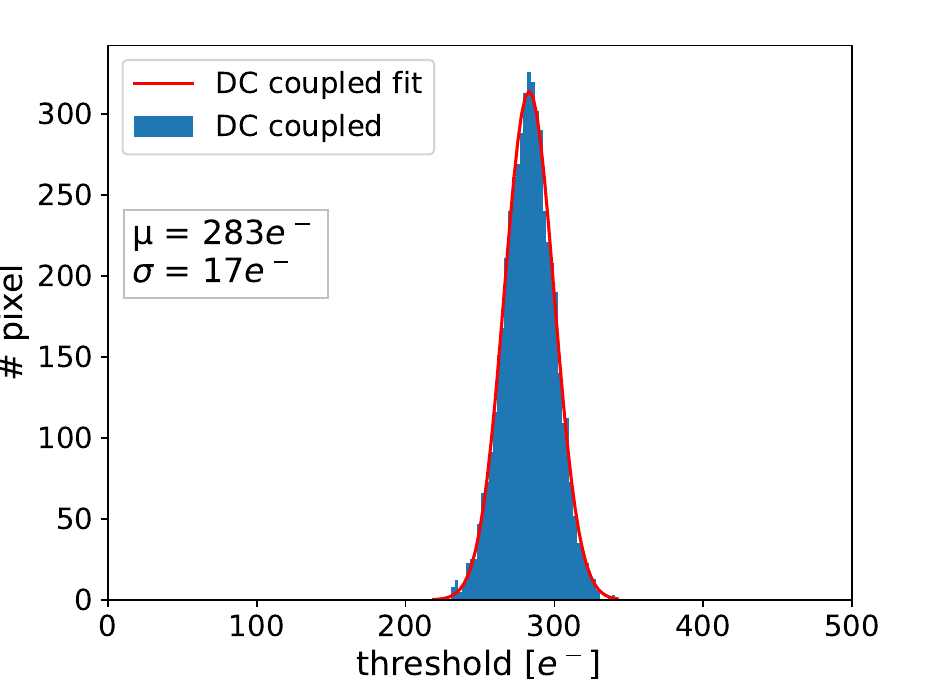}
        \caption{Threshold distribution of TJ-Monopix2 DC coupled FE with mean and dispersion for tuned settings.}
        \label{fig:threshold}
    \end{minipage}
\end{figure}
\begin{figure}
    \begin{minipage}[b]{0.49\textwidth}
\centering
\includegraphics[width=1\textwidth,  page=6 ,trim=0.2cm 0.cm 1.1cm .7cm, clip,]{images/20230602_000200_threshold_scan_interpreted.h5_Calibration_curve.pdf}
        \caption{Noise distribution of TJ-Monopix2 DC coupled FE with mean and dispersion for tuned settings.}
        \label{fig:noise}
    \end{minipage}%
    \hspace{0.15cm}
    \begin{minipage}[b]{0.49\textwidth}
\centering
\includegraphics[width=.98\textwidth]{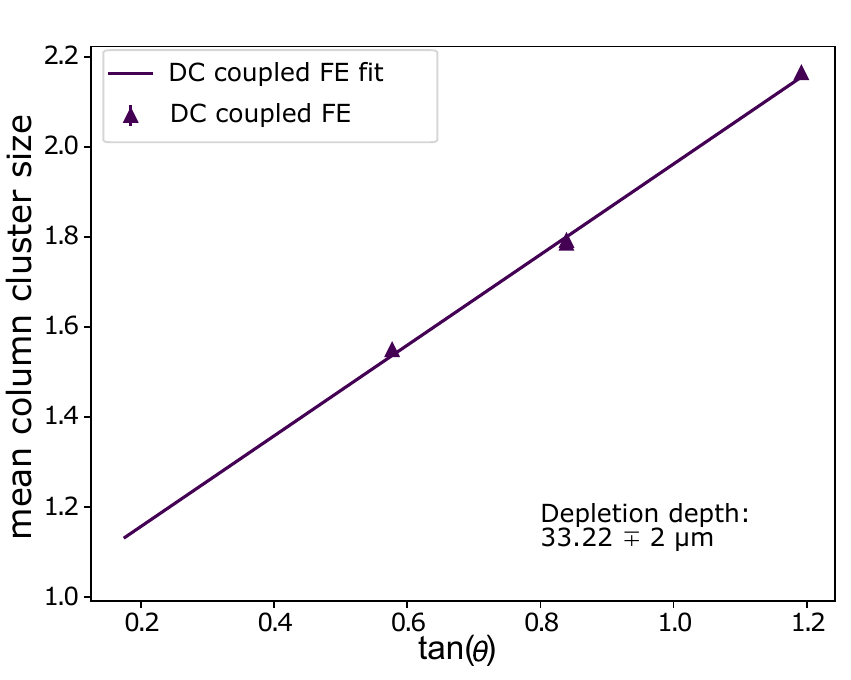}
        \caption{Depletion depth of TJ-Monopix2 DC coupled FE derived through a linear fit from angular measurements.}
        \label{fig:depl}
    \end{minipage}
\end{figure}
S-curve tests, employing the internal injection methods, are performed to measure the sensor threshold and noise. An example of an s-curve measurement is depicted in Figure \ref{fig:s-curve}, demonstrating distinct behaviour: minimal to no response until a certain minimal injected charge is reached, followed by a rapid and nearly linear increase in response up to a full occupancy of 1. Beyond this point, the curve reaches saturation. The threshold is determined by the injected charge at which a 50\% occupancy is reached. It can be adjusted both globally and locally, with local tuning achieved using a 3-bit TDAC available in each pixel. Since the threshold is determined individually for each pixel, the dispersion of thresholds across the sensor is illustrated in Figure \ref{fig:threshold}. For this example, a threshold of 283\;$e^-$ was measured, characterised by a standard deviation equivalent to  17\;$e^-$. The noise level, quantified by the slope of the s-curve, was found to be 8\;$e^-$, shown in Figure \ref{fig:noise}.

In June 2022, beam tests were conducted at DESY using an electron beam with energies in the range of 3-5\;GeV. The Mimosa EUDET-Telescope \cite{Jansen_2016} was used as a reference for the particle location. Notably, the chips employed for the tests were unirradiated. During these tests, preliminary settings were used, featuring high thresholds set at approximately 500\;$e^-$.

\begin{figure}
    \begin{minipage}[b]{0.49\textwidth}
\centering
\includegraphics[width=0.9\textwidth,trim= 0cm 0cm 0cm 0.7cm, clip]{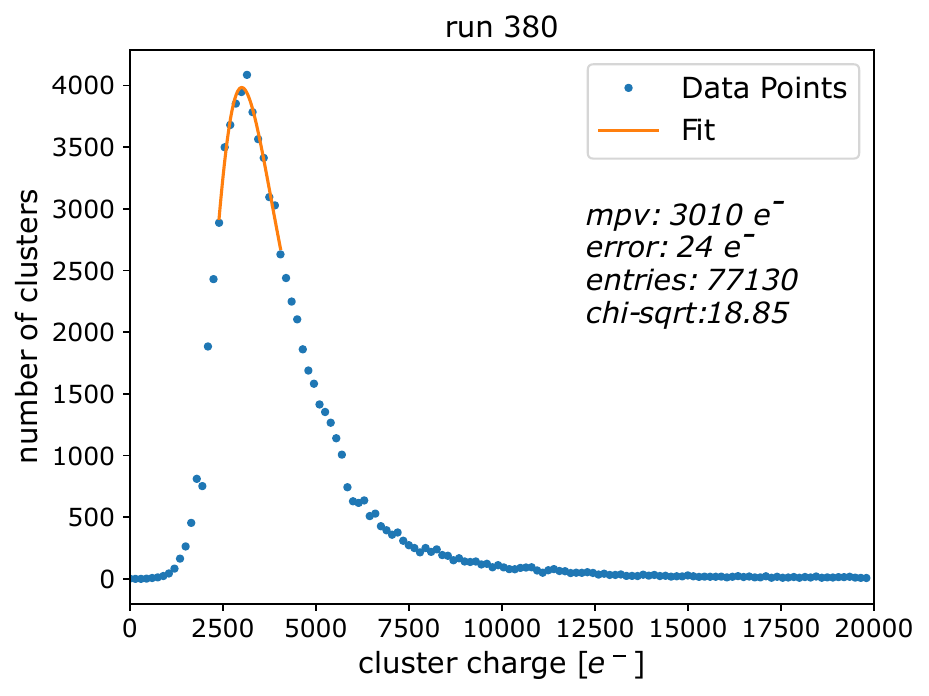}
        \caption{Cluster charge distribution for the beam measurement.}
        \label{fig:clustercharge}
    \end{minipage}
    \hspace{0.15cm}
    \begin{minipage}[b]{0.49\textwidth}
\centering
\includegraphics[width=1.1\textwidth, page=34]{images/Histos-TJ2-run3805V_calib_eta-run380-5V_calib_eta-reco-NF.pdf}
        \caption{$(99.54\pm0.04)\%$ Hit detection efficiency of an DC-coupled pixel flavour at 3\;V  bias voltage.}
        \label{fig:eff}
    \end{minipage}
\end{figure}

The cluster charge distribution plot, obtained from the electron beam measurements, provides insights into energy deposition patterns and is shown in Figure \ref{fig:clustercharge}. The MPV (Most Probable Value) of the cluster charge is determined through a Landau fit, where the signal of each pixel is corrected by a calibration factor, resulting in $(3010 \pm 24)\;e^-$. The derived depletion depth of approximately $(33\pm2)$\;\textmu m, as determined from an angular scan shown in Figure \ref{fig:depl}, aligns with the anticipated signals. The depletion depth estimation is based on geometry, linking cluster size, incident angle $\alpha$, and active depth $d$. The depletion depth is derived through a linear fit to mean cluster size concerning the tangent of the rotation angle. The calculated deposited energy is approximately $(3200\pm194)\;e^-$.

Efficiency and resolution are critical parameters under evaluation. The hit efficiency is defined as the ratio of matched hits to total tracks, denoted by $\epsilon = \frac{n_{\text{matched}}}{n_{\text{tracks}}}$. For a threshold of approximately 500\;$e^-$, the hit efficiency is at $99.54\% \pm0.04\%$, as shown in Figure \ref{fig:eff}. Also visible in the figure is the high level of homogeneity of the efficiency over the enabled sensor matrix.  In terms of spatial precision, at an incident energy of 4\;GeV and with perpendicular incident particles,  the cluster position resolution is achieved at around $9.15\;$\textmu m. This resolution surpasses the expected $\frac{\text{pitch}}{\sqrt{12}}$ value, which is approximately $9.5\;$\textmu m. The measured results fall within the requirements set in the beginning for the VTX upgrade.

The next assessment phase involves irradiation ranging from $10^{14}$ to $10^{15}$~n$_\text{eq}$/cm\textsuperscript{2}, and in July 2023, a test beam examined the sensor's response under these conditions. Additional sensor matrices were irradiated at the Ljubljana TRIGA reactor from the Jozef Stefan Institute at different fluences within the specified range relevant to VTX, but they have not been assembled yet.

\section{Conclusion and Outlook}
In conclusion, simulations indicate substantial improvements in the tracking and physics performance with a five layer CMOS pixel detector replacement for the VXD. The development and testing of the VTX is ongoing. Significant advancements have been achieved in the mechanical design of the VTX.

The careful characterisation of the TJ-Monopix2 sensor matrix is crucial for the VTX Upgrade proposal, since its sensor matrix design will be carried over to the development of the OBELIX detector. Its performance figures in the non-irradiated state have demonstrated the ability to meet the experiment's requirements. The successful test beam and stable module operation over extended periods provide confidence in the reliability and functionality of the detector. Beam measurements were performed during the time from July the 3rd to July the 6th 2023 at DESY on irradiated sensors, as well as additional measurements on unirradiated sensors, providing valuable insights for optimising the design of the upcoming OBELIX chip. The data analysis is currently ongoing and the OBELIX design is in development, with the aim of submitting it at the end of the year.

 The VTX collaboration will contribute to a conceptual design report for the Belle II upgrade in late 2023 and continue to work on testing and developing the system, aiming for an installation in 2027/2028.

In summary, the upgrade to Depleted CMOS MAPS technology offers a promising path for enhancing the SuperKEKB vertex detector, leading to improved performance and data quality in the high luminosity environment expected after long shutdown 2.
\acknowledgments
The measurements leading to these results have been performed at the Test Beam Facility at DESY Hamburg (Germany), a member of the Helmholtz Association (HGF). This work has received funding from the European Union’s Horizon 2020 Research and Innovation programme under Grant Agreements no 101004761 (AIDAinnova) and no 101057511 (EURO-LABS), was supported by Grant CIDEGENT/2018/020 of Generalitat Valenciana (Spain) and has been published under the framework of the IdEX University of Strasbourg.

\bibliographystyle{unsrt}
\bibliography{bib_vtx}

\end{document}